%% file: wprime-PRL.tex
\documentclass[aps,prl,twocolumn,showpacs,groupedaddress]{revtex4}  
\usepackage{graphicx}  
\usepackage{dcolumn}   
\usepackage{bm}        
\usepackage{amssymb}   

\newcommand{\singletop} {\sc singletop}
\newcommand{\pythia} {\sc pythia}
\newcommand{\geant} {\sc geant}
\newcommand{\alpgen} {\sc alpgen}
\begin{document}
\hspace{5.2in} \mbox{FERMILAB-PUB-08/067-E}
\title{Search for $W{^\prime}$ boson resonances decaying to a top quark and a
  bottom quark}
\input list_of_authors_r2.tex  
\date{March 21, 2008}

\begin{abstract}
We search for the production of a heavy $W{^\prime}$ gauge boson that decays
to  third generation quarks in 0.9 fb$^{-1}$ of $p\overline{p}$ collisions
at $\sqrt{s} = 1.96$~TeV, collected  with the D0 detector at the Fermilab
Tevatron collider. We find no significant excess in the final-state
invariant mass distribution and set upper limits on the  production cross
section times branching fraction. For a left-handed $W{^\prime}$ boson with
SM couplings,  we set a lower mass limit of 731~GeV. For right-handed
$W{^\prime}$ bosons, we set  lower mass limits of 739~GeV if the $W{^\prime}$
boson decays to both leptons and quarks and 768~GeV if the $W{^\prime}$ boson
decays only to quarks.  We also set limits on the coupling of the
$W{^\prime}$ boson to fermions as a  function of its mass.
\end{abstract}

\maketitle 

New massive charged gauge bosons, usually called $W{^\prime}$, are predicted
by various extensions of the standard model (SM). Noncommuting extended
technicolor, little Higgs, composite gauge bosons, grand unification, and
superstring theories  represent examples in which an extension of  the gauge
group leads to the appearance of a $W{^\prime}$ boson~\cite{Boos:2006xe}.

Direct searches for such $W{^\prime}$~bosons in leptonic final states
($\ell\nu$) lead to the  lower limit $M_{W{^\prime}}>1.0$~TeV~\cite{D0:2008},
assuming the $W{^\prime}$~boson couples  to fermions in the same way as the
SM $W$~boson. $W{^\prime}$~bosons that couple to  right-handed fermions may
not be able to decay to leptonic final states if  the corresponding
right-handed neutrinos are too massive. In this case, only  decays to
$q\overline{q}^\prime$ final states are possible, and the best limit, based
on decays of the $W{^\prime}$~boson to two light quark jets 
is $M_{W{^\prime}}>800$~GeV~\cite{D0:2004}.  
There are model-dependent upper limits on the mass of
$W{^\prime}$ bosons,  based on cosmological and astrophysical data, that
range from 549~GeV to  23~TeV~\cite{PDG:2006}.

In this Letter, we report a search for a $W{^\prime}$ boson that decays to
third  generation quarks ($W{^\prime} \to t\bar b$ or $\overline{t}{b}$).
For brevity, we will use the notation $tb$ to represent the
sum of the $t\bar b$ and the $\bar t {b}$ decay modes.

A $W{^\prime}$ boson that decays to $tb$ contributes to single top quark
production~\cite{Simmons:1996ws} for which evidence has been reported
recently~\cite{single-top-evidence}. Since the SM $W$~boson and a
hypothetical  $W{^\prime}$ boson with left-handed couplings both couple to
the same fermion multiplets, they interfere with each other. The interference
term may reduce the total rate by as much as (16--33)\%, depending on the
mass of the $W{^\prime}$ boson and its couplings~\cite{Boos:2006xe}.
Previous searches~\cite{CDF:2003} 
in this channel (neglecting interference effects) at
the Tevatron have led to the 95\% C.L. limits
$M_{W{^\prime}}>536$ GeV if the $W{^\prime}$ decays to $\ell\nu$ and to
$q\overline{q}^\prime$ and $M_{W{^\prime}}>566$ GeV if it only decays to
$q\overline{q}^\prime$.  A recent D0 analysis~\cite{D0wprime}, which takes
into account the interference, excludes masses between 200~GeV and 610~GeV
for a $W{^\prime}$~boson with left-handed SM-like couplings, between 200~GeV
and 630~GeV for a $W{^\prime}$~boson with right-handed couplings that decays
to $\ell\nu$ and $q\overline{q}^\prime$, and between 200~GeV and 670~GeV for
a $W{^\prime}$~boson with right-handed couplings that can only decay to
$q\overline{q}^\prime$.

The most general lowest-order effective Lagrangian for the interactions of a
$W{^\prime}$ boson with SM  fermions $f$ with generation indices
$i$ and $j$, is
$$ {\cal L} = \frac{V_{ij}}{2\sqrt{2}} g_w \overline{f}_i\gamma^\mu \bigl[
a^R_{ij} (1+{\gamma}^5) + a^L_{ij}(1-{\gamma}^5)\bigr]  W{^\prime}_\mu f_j +
\mathrm{h.c.},$$ 
where $V_{ij}$ is
the Cabibbo-Kobayashi-Maskawa~\cite{CKM} matrix element if the fermion  is a
quark, and $V_{ij}=\delta_{ij}$ if it is a lepton, $\delta_{ij}$ is the
Kronecker delta, $g_w$ is the weak coupling constant of the SM, and 
$a^L_{ij},\ a^R_{ij}$ are coefficients.
In this notation, $a^L_{ij}=1$ and $a^R_{ij}=0$ for a
so-called SM-like $W{^\prime}$~boson. This effective Lagrangian has been
incorporated into the {\sc comphep} package~\cite{Boos:2004kh} and used by
the {\singletop} event generator~\cite{Boos:2006af}. 
{\singletop} is used to
simulate SM single top quark production via the exchange of
a $W$ boson in the $s$- and $t$-channel, and the
$s$-channel $W{^\prime}$ signal, including interference with the SM 
$W$~boson. We simulate the complete chain of
$W{^\prime}$, top quark, and $W$~boson decays, taking into account finite
widths and all spin correlations between the production of resonance states
and their decay.  The top quark mass is set to 175~GeV, the CTEQ6L1 parton
distribution functions~\cite{CTEQ6L1} are used and the factorization scale is
set to $M_{W{^\prime}}$. Next-to-leading-order (NLO) corrections are included
in the {\singletop} generator, and normalization and matching between various
partonic subprocesses are performed such that not only the rates, but also
the shapes of distributions at NLO~\cite{Sullivan:2002jt}, are reproduced.

We generate samples of purely left-handed $W^\prime_L$ bosons with
$a^L_{ij}=1$ and $a^R_{ij}=0$, and purely right-handed $W^\prime_R$~bosons
with $a^L_{ij}=0$ and $a^R_{ij}=1$. $W^\prime_L$ bosons interfere with the
standard $W$~boson, but $W^\prime_R$ bosons couple to different final state
particles and therefore do not interfere with the standard $W$~boson. The
$\ell\nu$ decays of $W^\prime_R$~bosons involve a right-handed neutrino of
unknown mass, assumed to be $M_{\nu_R}>M_{W{^\prime}}$ 
or $M_{\nu_R}<M_{W{^\prime}}$. 
The $W{^\prime}$ width varies between 20~GeV and 30~GeV for
$W{^\prime}$~masses between 600~GeV and
900~GeV~\cite{Boos:2006xe,Sullivan:2002jt}. If $M_{\nu_R}>M_{W{^\prime}}$ and
only $q\overline{q^\prime}$ final states are open, the width is about 25\%
smaller. This does not have a significant effect on our search as the
experimental resolution for the $tb$ invariant mass  is much larger ($\approx$
90 GeV).  The branching fraction for $W{^\prime}\rightarrow tb$, is 
around 0.32 (0.24) for decays  only to quarks (quarks and leptons)
for a $W{^\prime}$ boson with a  mass of 700~GeV and varies slightly
with the mass. In the absence of
interference between $W$ and $W{^\prime}$~bosons, and if
$M_{\nu_R}<M_{W{^\prime}}$, there is no difference between $W^\prime_L$ and
$W^\prime_R$ for our search.  Since the current lower limit on the
mass of the $W^\prime$ boson is around 600~GeV~\cite{D0wprime}, 
we simulate $W^\prime_L$ and
$W^\prime_R$ bosons at seven mass values from 600 to 900~GeV to probe for
$W{^\prime}$ bosons with higher masses.

We analyze events with leptons, jets, and missing transverse momentum,
{\mbox{$\not\!\!p_T$}},  in the final state. The data were recorded by the D0
detector~\cite{d0det} between 2002 and 2005 using triggers that required a
jet and an electron or a muon. They correspond to~0.9~fb$^{-1}$ of integrated
luminosity. The event selection and trigger criteria are very similar to
those in  Ref.~\cite{single-top-evidence} and require exactly one isolated
electron (muon) with a momentum component transverse to the beam direction
$p_T > 15\ (20)$~GeV and pseudorapidity $|\eta| < 1.1\ (2.0)$,
$15<$ {\mbox{$\not\!\!p_T$}}  $<200$~GeV, a leading jet with
$p_T >25$~GeV and $|\eta|< 2.5$, and a second leading jet with $p_T >20$~GeV
and $|\eta| < 3.4$. We select events with two or three jets, counting all
jets with $p_T >15$~GeV and $|\eta| < 3.4$. Events with more than three jets
are excluded to reduce the $t\bar{t}$ background. Jets are reconstructed
using the Run II midpoint cone algorithm~\cite{cone} with cone size
$0.5$. Since we expect two $b$ quarks in the $W{^\prime}\rightarrow tb $
decay, we require at least one jet to be classified  (``tagged'') as a $b$
jet~\cite{btagalgo}. The data are divided into  eight independent channels
based on lepton flavor ($e$, $\mu$), jet multiplicity (2, 3), and number of
$b$-tagged jets (1, $\geq$2) to take into account the different signal
acceptances and signal-to-background ratios, which increases  the sensitivity
of the search.

Background yields are estimated using both Monte Carlo (MC) samples and data
in the same way as in Ref.~\cite{single-top-evidence}. Control data samples
are used to determine the multijet background from events in which a jet is
misidentified as an electron, or a muon from a semileptonic heavy flavor 
quark decay
is considered to be from the decay of a $W$~boson. The $t\bar t$ background
is estimated using the {\alpgen}~\cite{alpgen} MC event generator, 
normalized to the theoretical
cross section of $6.8\pm 1.2$~pb~\cite{ttxsec}. The $W$+jets background is
modeled using {\alpgen} and its yield is
normalized together with the multijet background, so that the total
background yield equals the observed number of events in data before
requiring a $b$-tagged jet. The fraction of $W$+jets events with heavy
flavors ($Wb\bar b$, $Wc\bar{c}$) is measured using data. In this way,  the
small contributions from $Z$+jets and diboson processes  ($WW$, $WZ$, $ZZ$)
are absorbed into the $W$+jets background normalization.  For the $W^\prime_R$
search, the SM single top quark production   is included in the
background. Because of their interference, the $s$-channel single top quark
production is considered part of the signal  for the $W^\prime_L$ search 
and only the $t$-channel single top quark production 
is included  in the background.
All parton-level MC samples are further processed with
{\pythia}~\cite{pythia} and a {\geant}~\cite{geant}-based simulation of the
D0 detector. Lepton and jet energies are corrected to reproduce the resolutions
observed in data.

The distinguishing feature of a $W{^\prime}$ signal is a  resonance structure
in the $tb$ invariant mass. However, we cannot directly measure the $tb$
invariant mass. Instead we reconstruct the invariant mass $\sqrt{\hat{s}}$ of
the leading two jets, the charged lepton, and the neutrino by adding their
measured momentum four-vectors. The missing transverse momentum is used to
obtain the $x$ and $y$-components of the neutrino momentum and the
$z$-component is the smaller of the two  $|p_z^{\nu}|$ 
values that makes the $\ell\nu$ mass
equal the $W$~boson mass.

\begin{figure}
\begin{center}
\includegraphics[scale=0.4]{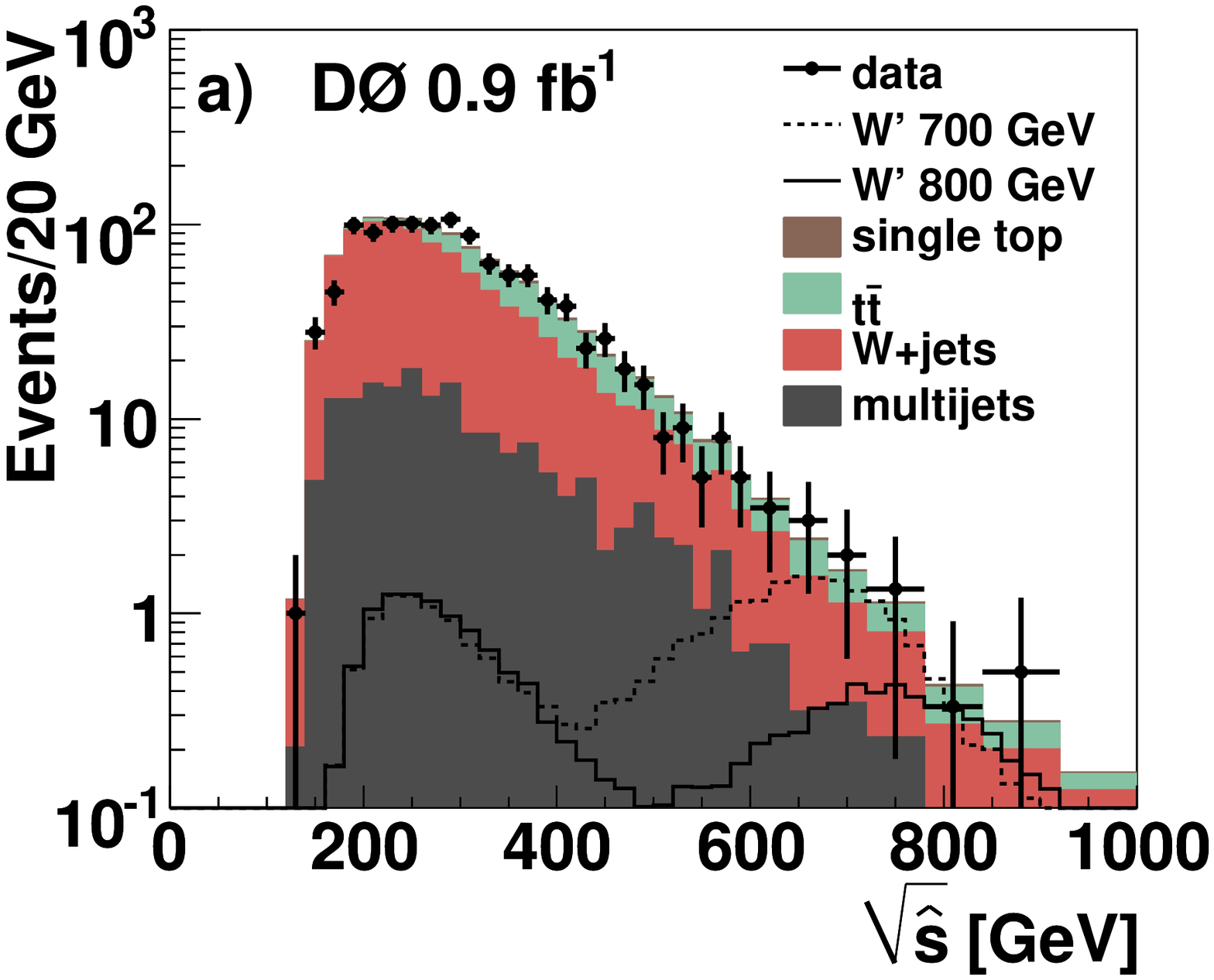}
\includegraphics[scale=0.4]{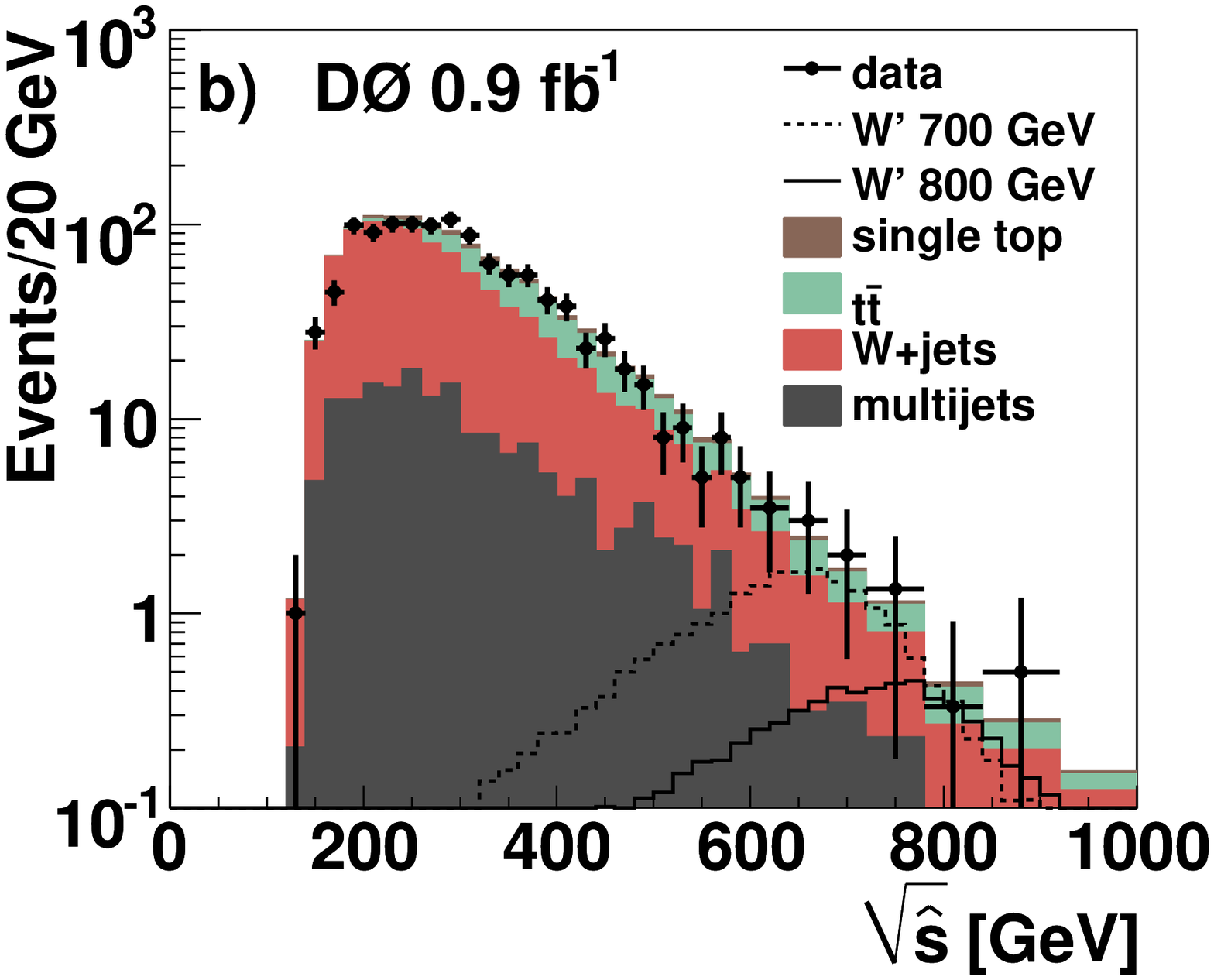}
\end{center}
\caption{$\sqrt{\hat s}$ distributions for the data and the SM
  background. Plot (a) shows the $W^\prime_L$ signal and plot (b) shows the
  $W^\prime_R$ signal at two different masses, normalized to the 
  NLO prediction (Table~\ref{tab:xs}). Events in the 
  eight subsamples (electron, muon, two and three jets, 
  single-tagged and double-tagged) are combined.
\label{fig:Wprime-mass-data}}
\end{figure}

The observed $\sqrt{\hat s}$ distribution in the data is consistent with the
background prediction within uncertainties and shows no evidence for a
signal (see Fig.~\ref{fig:Wprime-mass-data}).  Since we search for
$W^\prime$ bosons with masses greater than 600~GeV, we set upper limits  on
the $W{^\prime}$ boson production cross section times branching fraction 
to the $tb$ final state, $\sigma(p\bar{p}\rightarrow W{^\prime}) \times 
{\rm B}(W{^\prime} \rightarrow tb)$,
using the high tail of the $\sqrt{\hat{s}}$
distribution. 
Table~\ref{tab:xb} gives the  observed number of data events
and the expected background yields  for events with $\sqrt{\hat s}>400$~GeV
(chosen to improve the signal to background ratio).
The signal yields
corresponding to the same selection are listed in Table~\ref{tab:xs}.  We use
a Bayesian method~\cite{top-stat} with a flat nonnegative prior for the
signal cross section. The limits are derived  using a binned likelihood
constructed from the $\sqrt{\hat{s}}$ spectrum above 400~GeV,
taking into account all systematic uncertainties, and their correlations.
We also compute expected upper limits as
a measure of the sensitivity of the analysis. We combine the eight
independent subsamples to obtain the  limits listed in Table~\ref{tab:xs}.

\begin{table}[!h!tbp]
\begin{center}
\caption{Data and  SM background event yields.}\label{tab:xb}
\begin{tabular}{l  r@{ $\pm$ }l  r@{ $\pm$ }l }
\hline\hline
Process &\multicolumn{4}{c}{ Events }\\
        & \multicolumn{2}{c}{SM+$W^{\prime}_L$} search &  
\multicolumn{2}{c}{$W^\prime_R$ search}    \\
\hline
Single top                     & $6.4$ & $1.4$ & $10.2$ & $2.2$ \\
$t\bar{t}$                     & \multicolumn{4}{c} {$59.1\pm14.4$} \\
$W$+jets                       & \multicolumn{4}{c} {$91.0\pm 18.8$} \\
Multijets                      & \multicolumn{4}{c} {$29.7\pm\hphantom{0}5.9$} \\ 
Total background               &  $186.1$& $40.4$ &  $190.0$& $41.2$ \\
\hline
Data                           &  \multicolumn{4}{c}{182} \\
\hline\hline
\end{tabular}
\end{center}
\end{table}

In the evaluation of the systematic uncertainties,  we take into account the
uncertainties  in both the background normalization and the shape of the
$\sqrt{\hat{s}}$ distribution. Uncertainties in the integrated luminosity
(6.1\%), theoretical cross sections [(15--18)\%], branching fractions (1\%),
object identification efficiencies [(1--7.5)\%], trigger efficiencies
[(3--6)\%], jet fragmentation modeling (5\%), and the uncertainty in the
fraction of $W$+jets events with heavy quarks affect only the
normalization~\cite{single-top-evidence}. Uncertainties in the $b$-jet
simulation [(12--17)\%] and the jet energy scale calibration  [(1--20)\%]
affect both shape and normalization. Ranges represent the variations 
among the eight subsamples. 

\begin{table}
\caption{\label{tab:xs} NLO production cross sections $\times$ branching
  fraction to $tb$ in pb (theory), expected signal event yields (evts), and  
expected (exp) and observed (obs) 95\% C.L. upper limits for 
$\sigma(p\bar{p}\rightarrow W{^\prime})
\times {\rm B}(W{^\prime} \rightarrow tb)$ in pb.  
Theory I (II) corresponds to the case $M_{\nu_R}<M_{W{^\prime}}$ 
($M_{\nu_R}>M_{W{^\prime}}$).
The uncertainty on signal yields are around 20\%.
 }
\begin{center}
\begin{tabular}{lcccclrccc} \hline \hline
$M_{W{^\prime}}$ & \multicolumn{4}{c}{$W^\prime_L$} 
&  \multicolumn{5}{c} {$W^\prime_R$}  \\
 (GeV)   & Theory &  Evts &\ Exp\ &\ Obs\ \ \ &  \multicolumn{2}{c}{Theory} & Evts &
\ Exp\ &\ Obs\ \\
    &  &   &  & &\ \ (I)\  &\ (II)\ \ & & & \\\hline
600  &  2.17 & 58 &   0.69  &  0.66 & 2.10 & 2.79  & 61 &  0.67  &  0.58  \\
650  &  1.43 & 33 &   0.65  &  0.69 & 1.25 & 1.65  & 35 &  0.55  &  0.59  \\
700  &  1.01 & 19 &   0.69  &  0.74 & 0.74 & 0.97  & 20 &  0.50  &  0.54  \\
750  &  0.76 & 11 &   0.80  &  0.93 & 0.44 & 0.57  & 12 &  0.44  &  0.50  \\
800  &  0.62 & 6  &   1.04  &  1.23 & 0.26 & 0.34  & 7  &  0.42  &  0.47  \\
850  &  0.55 & 4  &   1.46  &  1.77 & 0.16 & 0.20  & 4  &  0.42  &  0.48  \\
900  &  0.51 & 3  &   2.35  &  2.79 & 0.09 & 0.12  & 2  &  0.40  &  0.44  \\
\hline \hline
\end{tabular}
\end{center}
\end{table}

\begin{figure*}[ht]
\begin{center}
\setlength{\unitlength}{1.0cm}
\begin{picture}(18.0,5.0)
\put(0.1,0.2){\includegraphics[scale=0.32]{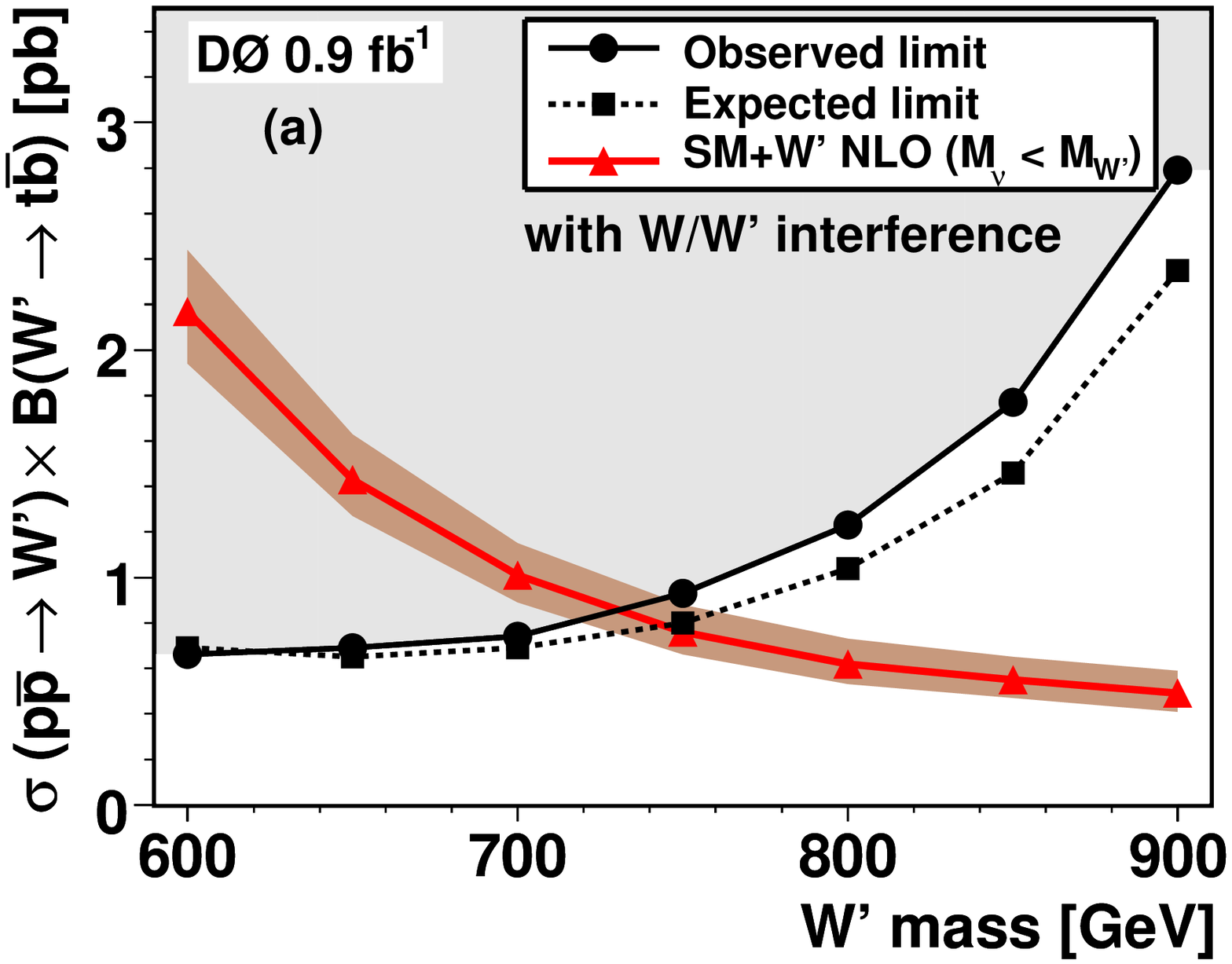}}
\put(6.1,0.2){\includegraphics[scale=0.32]{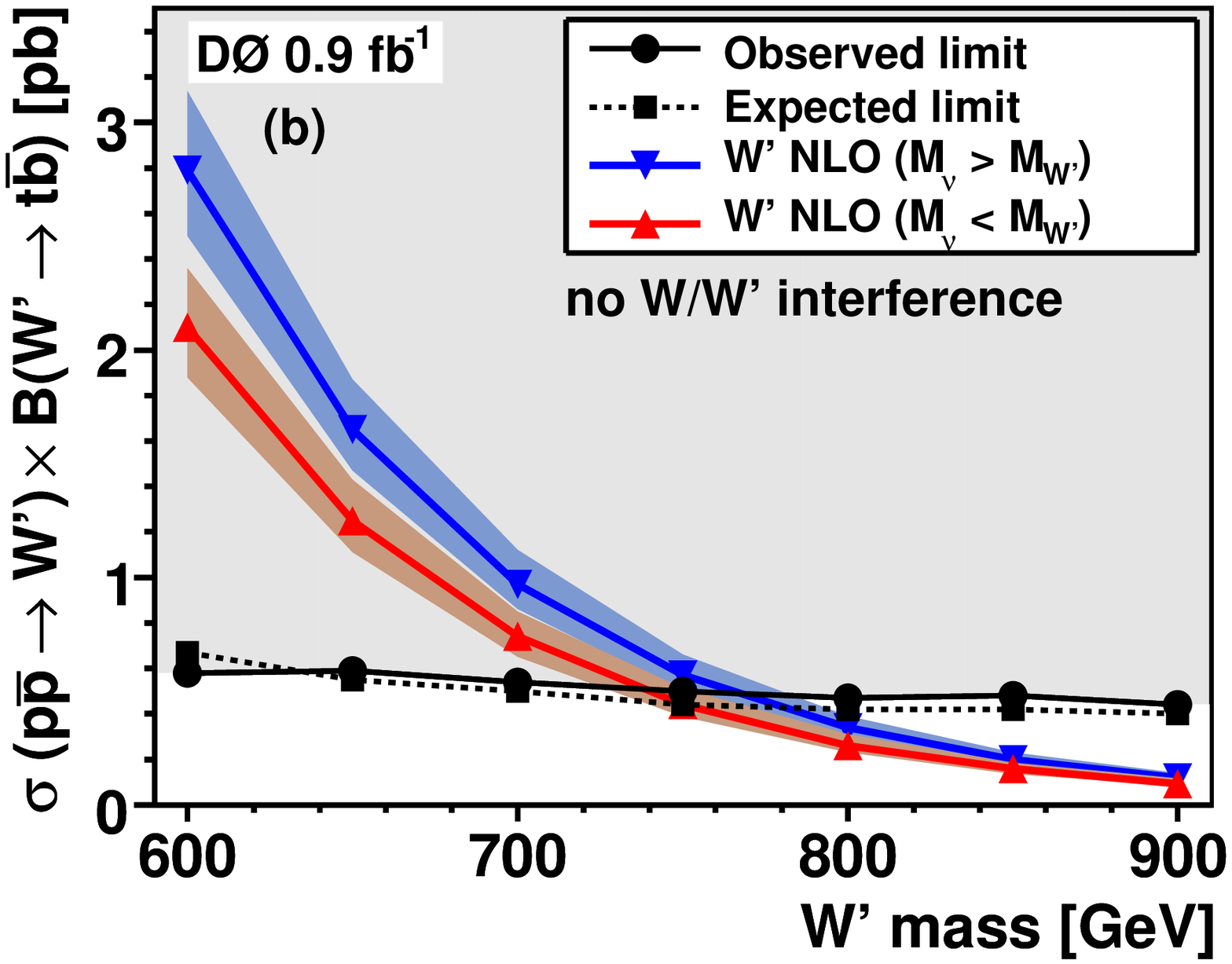}}
\put(12.1,0.2){\includegraphics[scale=0.32]{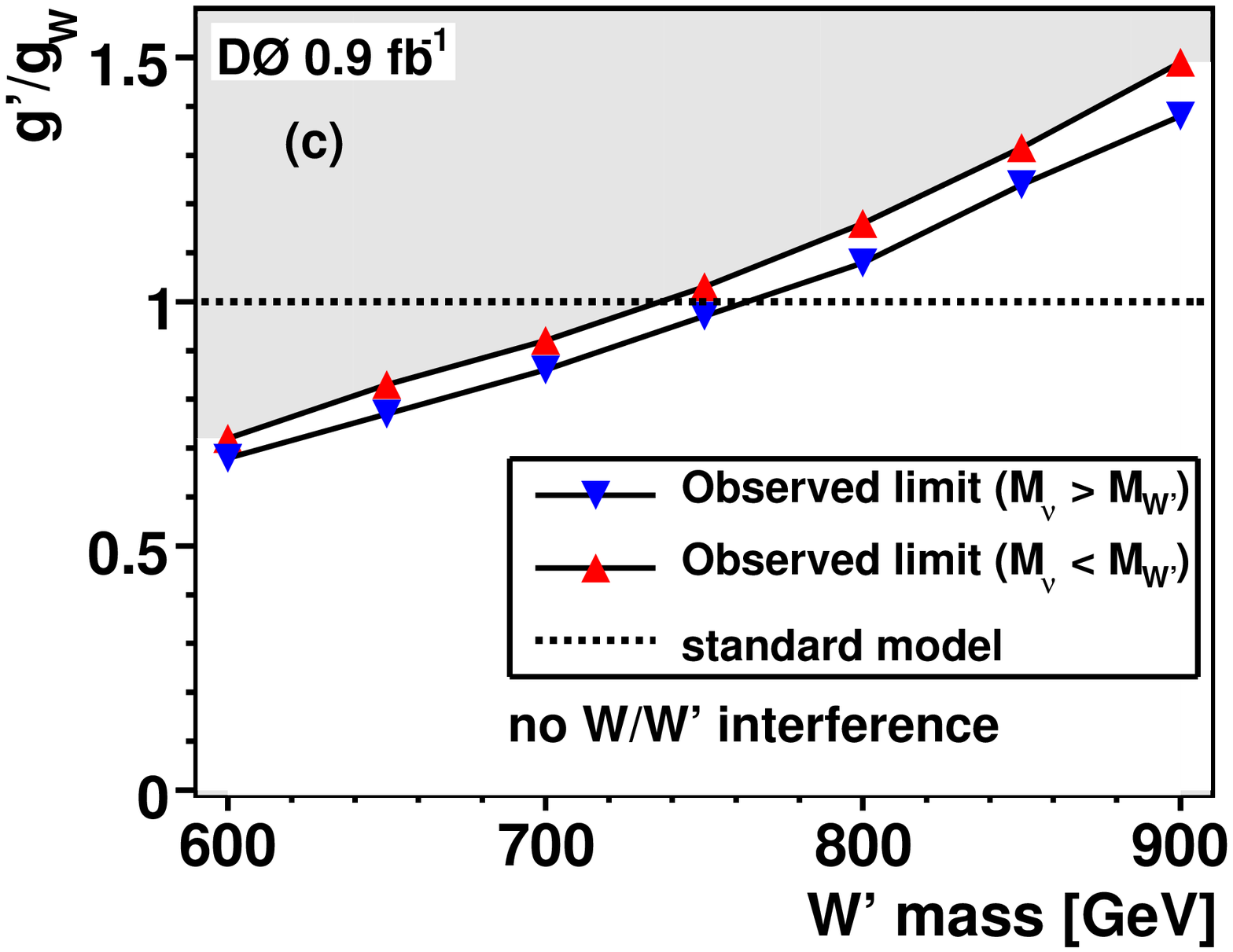}}
\end{picture}
\caption{NLO theory cross sections and 95\% C.L. limits for 
$\sigma_{W{^\prime}} \times {\rm B}(W{^\prime} \rightarrow tb)$
as a function of  $W{^\prime}$ mass for (a) $W^\prime_L$ production and (b) 
$W^\prime_R$ production. Observed limits on the ratio of coupling 
constants $g^\prime/g_w$ are shown in (c). 
The shaded regions are excluded by this analysis.
 \label{fig:limit}}
\end{center}
\end{figure*}

The observed 95\% C.L. upper limit of $\sigma(p\bar{p}\to W{^\prime}) \times
{\rm B}(W{^\prime}\rightarrow tb)$  compared to the NLO theory predictions
are shown in Fig.~\ref{fig:limit}   for (a) $W^\prime_L$ and (b) $W^\prime_R$
production cross sections. For the $W^\prime_L$~boson, we show the total
cross section for $s$-channel single top quark production including the SM
diagram, the $W{^\prime}$ diagram, and their
interference~\cite{Boos:2006xe}. In this case the limit applies to the
total $s$-channel single top production. 
The $k$-factors needed to scale the
$W^\prime_L$ cross section to NLO, the NLO cross sections for the
$W^\prime_R$~boson, and the expected theoretical uncertainty are taken from
Ref.~\cite{Sullivan:2002jt}. Using the nominal (${\rm nominal}-1\sigma$)
values of the   theoretical cross section,  the lower limit for
$W^\prime_L$~mass is  731 (718)~GeV. For  $W^\prime_R$~bosons that decay
only  to $q\overline{q}$ the limit is  768 (750)~GeV; 739 (725)~GeV if the
leptonic decay is also allowed.

Limits for the gauge couplings $g^\prime=g_w a_{ij}^L$ or $g^\prime=g_w
a_{ij}^R$, depending on the model, of the $W{^\prime}$~boson can be derived
from the cross section limits. Since the leading-order $s$-channel production
diagram has two $W{^\prime} q \overline {q}^\prime$ vertices,
$\sigma(p\bar{p}\to W{^\prime}) \times {\rm B}(W{^\prime}\rightarrow tb)$ is
proportional to  $g{^\prime}^4$.  Figure~\ref{fig:limit}(c) shows the
observed limit for $g^\prime/g_w$. We exclude gauge couplings above 
0.68 (0.72)~ $g_w$ for $W{^\prime}$~bosons with a mass of  600~GeV for the case
$M_{\nu_R}>M_{W{^\prime}}$  ($M_{\nu_R}<M_{W{^\prime}}$).

We have performed a search for $W{^\prime}$~bosons that decay to $tb$, using
0.9~fb$^{-1}$ of data recorded by the D0 detector. We find no evidence for
$W{^\prime}$~boson production and set 95\% C.L. upper limits on 
$\sigma(p\bar{p}\rightarrow W{^\prime}) \times 
{\rm B}(W{^\prime} \rightarrow tb)$. We use the nominal
value of the theoretical cross section to set limits on the mass of 
the $W^\prime$ bosons. We exclude
$W^\prime_L$~bosons with left-handed, SM-like couplings with masses below
731~GeV. For $W^\prime_R$ bosons with right-handed couplings, we set a lower
mass limit of 739~GeV if the $W{^\prime}$~boson can decay to leptons and to
quarks. If the $W{^\prime}$ decays only to quarks, the lower mass limit is
768~GeV. We also constrain the $W{^\prime}$ gauge coupling and
exclude couplings above 0.68 (0.72)~$g_w$ for $W^\prime$~bosons with a mass
of 600~GeV that only decay to quarks (leptons and quarks).  These limits
represent a significant improvement over previously published
results~\cite{CDF:2003, D0wprime}.

\input acknowledgement_paragraph_r2.tex   

\end{document}

%% file: list_of_authors_r2.tex
%
\author{V.M.~Abazov$^{36}$}
\author{B.~Abbott$^{75}$}
\author{M.~Abolins$^{65}$}
\author{B.S.~Acharya$^{29}$}
\author{M.~Adams$^{51}$}
\author{T.~Adams$^{49}$}
\author{E.~Aguilo$^{6}$}
\author{S.H.~Ahn$^{31}$}
\author{M.~Ahsan$^{59}$}
\author{G.D.~Alexeev$^{36}$}
% ...

\affiliation{\vspace{0.1 in}(The D\O\ Collaboration)\vspace{0.1 in}}
% ...
\affiliation{$^{6}$University of Alberta, Edmonton, Alberta, Canada,
                Simon Fraser University, Burnaby, British Columbia, Canada,
                York University, Toronto, Ontario, Canada, and
                McGill University, Montreal, Quebec, Canada}
% ...
\affiliation{$^{29}$Tata Institute of Fundamental Research, Mumbai, India}
% ...
\affiliation{$^{31}$Korea Detector Laboratory, Korea University, Seoul, Korea}
% ...
\affiliation{$^{36}$Joint Institute for Nuclear Research, Dubna, Russia}
% ...
\affiliation{$^{49}$Florida State University, Tallahassee, Florida 32306, USA}
% ...
\affiliation{$^{51}$University of Illinois at Chicago,
                Chicago, Illinois 60607, USA}
% ...
\affiliation{$^{59}$Kansas State University, Manhattan, Kansas 66506, USA}
% ...
\affiliation{$^{65}$Michigan State University,
                East Lansing, Michigan 48824, USA}
% ...
\affiliation{$^{75}$University of Oklahoma, Norman, Oklahoma 73019, USA}
% ...

%% file: acknowledgement_paragraph_r2.tex
%
We thank the staffs at Fermilab and collaborating institutions, 
and acknowledge support from the 
DOE and NSF (USA);
CEA and CNRS/IN2P3 (France);
FASI, Rosatom and RFBR (Russia);
CNPq, FAPERJ, FAPESP and FUNDUNESP (Brazil);
DAE and DST (India);
Colciencias (Colombia);
CONACyT (Mexico);
KRF and KOSEF (Korea);
CONICET and UBACyT (Argentina);
FOM (The Netherlands);
STFC (United Kingdom);
MSMT and GACR (Czech Republic);
CRC Program, CFI, NSERC and WestGrid Project (Canada);
BMBF and DFG (Germany);
SFI (Ireland);
The Swedish Research Council (Sweden);
CAS and CNSF (China);
and the
Alexander von Humboldt Foundation.